# Super-resolution photoacoustic imaging through a scattering wall

**Donald B. Conkey[1†], Antonio M. Caravaca-Aguirre[1†], Jacob D. Dove[2†], Hengyi Ju[2], Todd W. Murray[2], and Rafael Piestun[1]***

1. Dept. of Electrical, Computer, and Energy Engineering, University of Colorado, Boulder, Colorado, 80309
2. Dept. of Mechanical Engineering, University of Colorado, Boulder, Colorado, 80309

**\*** piestun@colorado.edu

[†] These authors contributed equally to this work

**Imaging through opaque, highly scattering walls is a long sought after capability with potential applications in a variety of fields[1]. The use of wavefront shaping to compensate for scattering has brought a renewed interest as a potential solution to this problem[2–9]. A key to the practicality of any imaging technique is the capability to focus light without direct access behind the scattering wall. Here, we address this problem using photoacoustic feedback for wavefront optimization. By combining the spatially non-uniform sensitivity of the ultrasound transducer to the generated photoacoustic waves with an evolutionary competition among optical modes, the speckle field develops a single, high intensity focus significantly smaller than the acoustic focus used for feedback. Notably, this method is not limited by the size of the absorber to form a sub-acoustic optical focus. We demonstrate imaging behind a scattering medium with up to ten times improvement in signal-to-noise ratio (SNR) and five to six times sub-acoustic resolution.**

Recent advances in wavefront shaping have made imaging through scattering walls a possibility[7–12]. By precompensating the optical wavefront, the light propagation can be controlled through and beyond scattering materials[2,3,13,14]. Most existing techniques, however, are limited by their need to generate feedback from behind or inside a scatterer with direct invasive access. Several recent techniques have overcome this limitation using acoustic waves, either with an ultrasound guide-star[15] or a photoacoustic feedback[16–18]. Two modifications of the ultrasound guide-star method have allowed for the creation of an optical focus smaller than the acoustic spot size [19,20]. Si et al.[19] frequency modulated diffuse light with a focused ultrasound transducer and then time reversed the encoded light using optical phase conjugation. Through iterative time reversal, a convergence of the light to the central region of the ultrasound focal volume where the modulation is the strongest was observed, resulting in an increase in the optical intensity in a sub-acoustic focal region. Alternatively, the time reversal of variance encoded light (TROVE) approach was also shown to break the acoustic resolution barrier by isolating the spatial location of optical speckles within the ultrasound focus. While both of these techniques are promising for deep fluorescence imaging in scattering materials, their inherent low SNR could limit implementation

Photoacoustics, where acoustic waves are generated from optical absorption and subsequent thermal expansion, have also been used for wavefront optimization[16,17]. In this case, the pressure of the generated acoustic wave is proportional to the light fluence. Hence, the detected photoacoustic wave provides a measure of the local light intensity, facilitating feedback and allowing for wavefront optimization. Only lately has imaging through scattering media been demonstrated using photoacoustic feedback[18]. Without optical focusing, photoacoustic resolution is determined by the ultrasound frequency and numerical aperture of the transducer. Considering a focused ultrasound transducer, uniform optical absorption excites acoustic waves throughout the focal region. However, as shown below, introduction of a scattering material creates a non-uniform speckle field with separate and distinct speckle grains coherently combining to produce the photoacoustic response. Previous work using photoacoustic feedback with wavefront shaping has assumed that this feedback would create an optical focus limited by the size of the acoustic focus[16,17]. With this assumption, a smaller optical focus could be created only when optimizing with sub-acoustic sized absorbers as targets[17].

In this communication we demonstrate that by introducing a scattering element into the optical illumination path both SNR and resolution can be improved with wavefront shaping, enabling super-resolution imaging. To accomplish this, the optical intensity of the speckle field within the transducer focus is redistributed to create a single overpowering speckle smaller than the ultrasound transducer focus. Thus, the speckle acts as an optical focus and improves both photoacoustic emission and resolution, effectively allowing optical resolution photoacoustic microscopy[21] deep through scattering media. Such capabilities enable superb three-dimensional (3D) imaging of complex biological structures.

Optimizing a speckle field within the ultrasound transducer focal region presents significant challenges, as several speckle modes are present (Fig. 1a and b). If multiple speckles contribute with equal weighting the feedback signal uniformly mixes the information coming from individual speckles and does not lead to focus optimization. However, the transducer focal region does not detect acoustic waves uniformly. Instead, a focused ultrasound transducer displays a Gaussian spatial sensitivity, which essentially weights the speckle field; giving higher weighting to speckles closer to the center (Fig. 1b). By preferentially weighting a single speckle mode, the spatially non-uniform photoacoustic feedback localizes the optical intensity to a single speckle smaller than the acoustic focus (Fig. 1a and b).

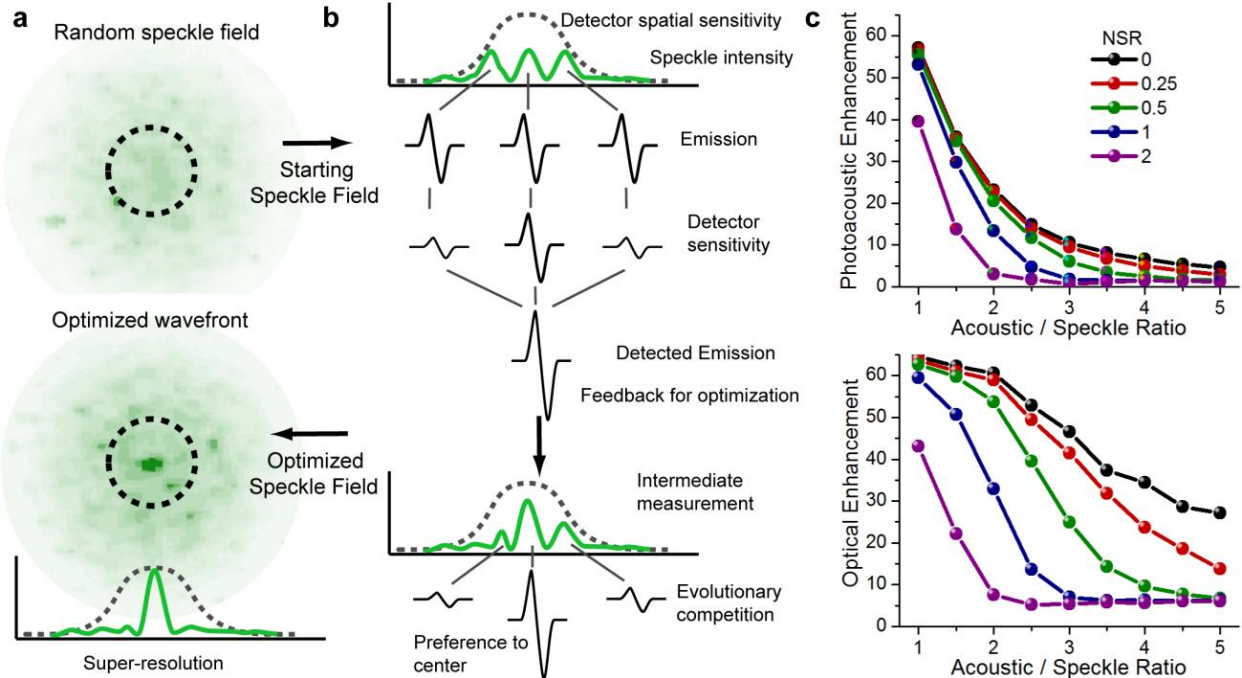

**Figure 1 | Optical focus creation with photoacoustic feedback: illustration and simulation. a**, The transducer focal region (dotted circle) illuminated with a speckle field (top) and the optimized speckle field with a sub-acoustic optical focus (bottom). **b**, Profile of speckle grains and the detector spatial sensitivity. Each photoacoustic emission from the speckle grains is weighted by the Gaussian spatial sensitivity of the acoustic transducer. These emissions coherently combine to provide the photoacoustic feedback signal. During wavefront optimization with the photoacoustic feedback an optical focus forms at the center of the acoustic transducer due to the weight preference. **c**, Simulation results of the photoacoustic and optical enhancements achieved after genetic algorithm optimization with various acoustic focus to speckle size ratios and noise-to-signal ratios (NSR).

Simulations of the wavefront optimization elucidate the mechanism by which a spatially non-uniform sensitivity produces a tight optical focus when multiple modes participate in the feedback (see Supplementary Information for methods). Here, we use an evolutionary algorithm[22] that creates competition among modes within the acoustic focus. The number of modes, determined by the acoustic spot to optical speckle size ratio, and noise (noted as noise-to-signal ratio: NSR) are varied and the optimized photoacoustic signal and optical field analyzed. The wavefront optimization feedback uses an equivalent photoacoustic signal by integrating the total weighted light fluence throughout the acoustic focus. The signal improvement is assessed by photoacoustic enhancement, defined as the optimized photoacoustic signal divided by the mean of the photoacoustic signals of the initial population in the genetic algorithm[22]. In contrast, the optical enhancement evaluates the improvement in the maximum intensity of an individual speckle output mode as the ratio between its intensity and the initial average intensity. Interestingly, the optical enhancement outperforms the photoacoustic enhancement as the number of modes is increased (Fig. 1c). The photoacoustic enhancement is integrated over the acoustic focal area while the optical enhancement is not. This indicates the optical energy localizes to an area smaller than the acoustic focus, creating an optical focus regardless of the small speckle size compared to the acoustic feedback. Furthermore, the optimized optical field most likely positions the enhanced speckle at the center of the acoustic focus. Thus, despite optimizing the wavefront based on feedback from multiple speckle modes within the acoustic focus, the light localizes to a single speckle, creating an optical focus.

We experimentally demonstrate the ability to localize a scattered light field to a sub-acoustic focus via photoacoustic feedback using the system shown in Fig. 2a. For this demonstration a 50 μm diameter black alpaca hair is selected for photoacoustic feedback, because it overfills both the acoustic transducer focal region and the speckle spot size (13 μm as measured through autocorrelation of Fig. 2b). The high bandwidth acoustic transducer provides a wide range of frequency feedback. Because the transducer focal size inversely relates to frequency, the photoacoustic optimization signal is high-pass filtered, effectively limiting the acoustic focal diameter (-6 dB) to 38 μm, as shown by the dashed line in Fig. 2d. After genetic algorithm optimization of the optical wavefront through phase modulation with the SLM, the photoacoustic signal is enhanced by 8.5. The optimization creates a 13 μm (FWHM) focused spot (Fig. 2c and e), with an optical enhancement of 24, or three times the photoacoustic enhancement. As predicted, a small, single speckle optical focus is created despite integrating many speckle output modes within the photoacoustic feedback signal.

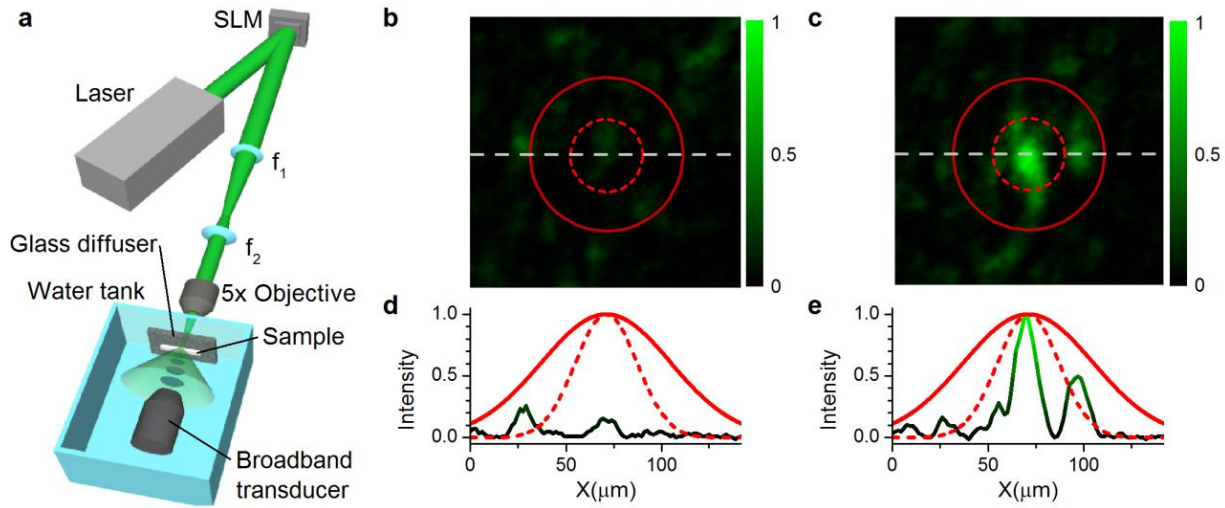

**Figure 2 | Experimental sub-acoustic focus creation. a**, Experimental set-up. **b**, The speckle field without the optimized wavefront. **c**, The sub-acoustic optical focus generated by the optimized wavefront. The red circles show the approximate filtered transducer focal region (80 MHz, -6 dB, dashed line) and focal spot size at the frequency peak of the detected photoacoustic response (50 MHz, -6 dB, solid line). **d,e** Cross sections of the approximate transducer focal regions (red) and the corresponding optical intensity distribution (black).

Next, we demonstrate the resolution improvement by imaging a black alpaca hair (25 μm diameter) placed behind a scatterer. To understand the nature of the resolution improvement, we perform photoacoustic imaging by scanning the hair under three different optical fields (uniform, random speckle and optimized speckle). A photoacoustic image of the hair illuminated with a uniform optical field is shown in Fig. 3a. Here, the measured photoacoustic FWHM of the hair is found to be on average 79 μm, which agrees with the expected size given the acoustic resolution (see Supplementary Information). With a glass diffuser placed in the optical path a random speckle field is created. The photoacoustic image reconstructed with this illumination is shown in Fig 3b. Interestingly, the FWHM of the hair decreases to an average of 55 μm. The random speckle field does show a resolution improvement, likely due to the speckle arrangement within the transducer (i.e. a few intense speckles that provided improved resolution). Through genetic algorithm optimization of the optical wavefront on a large (43 μm diameter) alpaca hair the photoacoustic signal is enhanced by 6. Photoacoustic imaging with the optimized wavefront provides both the highest SNR and the best resolution as shown in Fig 3c. The measured FWHM of the alpaca hair decreases to 30 μm. Assuming the optimization creates an optical focus of similar size to that of Fig. 2c, the resolution increases by five to six times compared to the acoustic resolution.

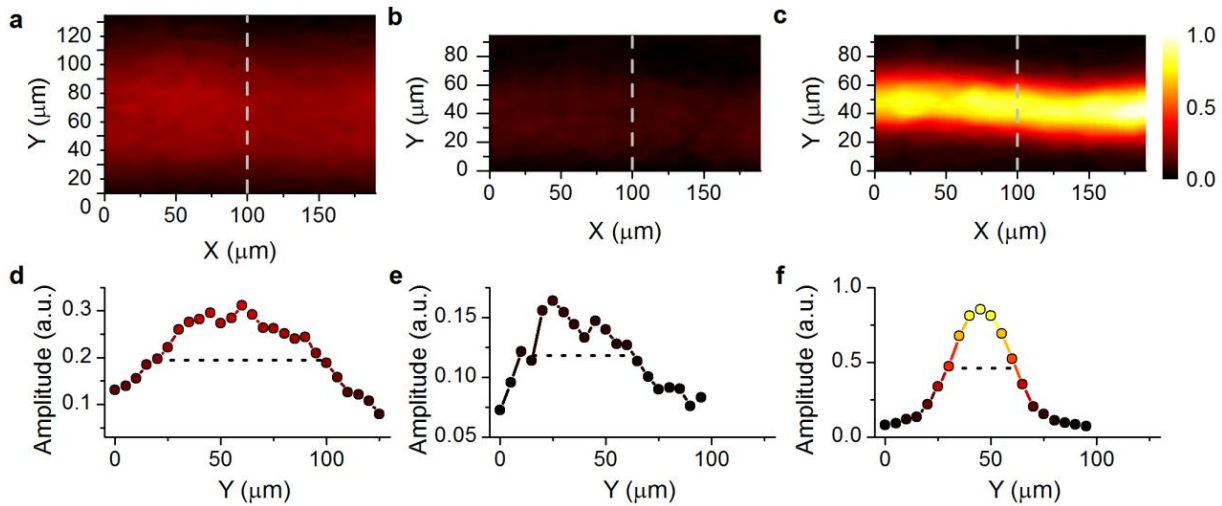

**Figure 3 | Resolution measurement with 25 μm black alpaca hair.** The photoacoustic images created with different optical illuminations: **a**, acoustic resolution using uniform illumination (no scatterer), **b**, random optical speckle field from scatterer, **c**, wavefront optimized speckle focus. **d**, **e**, and **f** show cross sections corresponding to the white dashed lines.

Super-resolution photoacoustic imaging performed on a sweat bee wing, which provides an interesting complex structure, reveals significant resolution and signal improvement. As with the alpaca hair, we present photoacoustic images of the wing with uniform (Fig. 4a) and random speckle (Fig. 4b) illuminations for comparison. For the wavefront optimization we select an extended feedback region (~40x40 μm) at the intersection of the large vein on the leading edge of the wing and a branching vein. After optimization the signal strength improved by a factor of 10 over the random speckle field illumination. We image the bee wing by scanning it behind the scatterer (Fig. 4c). The photoacoustic image significantly improves after optimization both in resolution and SNR, which facilitates 3D image reconstruction by sectioning the photoacoustic waveform temporally[18] (see movie). With the uniform and random speckle fields small features, visible in the optical image (Fig. 4d), such as hairs spaced 15-30 μm apart on the wing result in a high background signal and broadening of the prominent vein features. However, with the optimized focus image, the hairs are individually resolvable (see Supplementary Information). By treating the 4 μm wide hairs as point sources we measure the optical resolution to be 13.2 μm through autocorrelation of the imaged hairs.

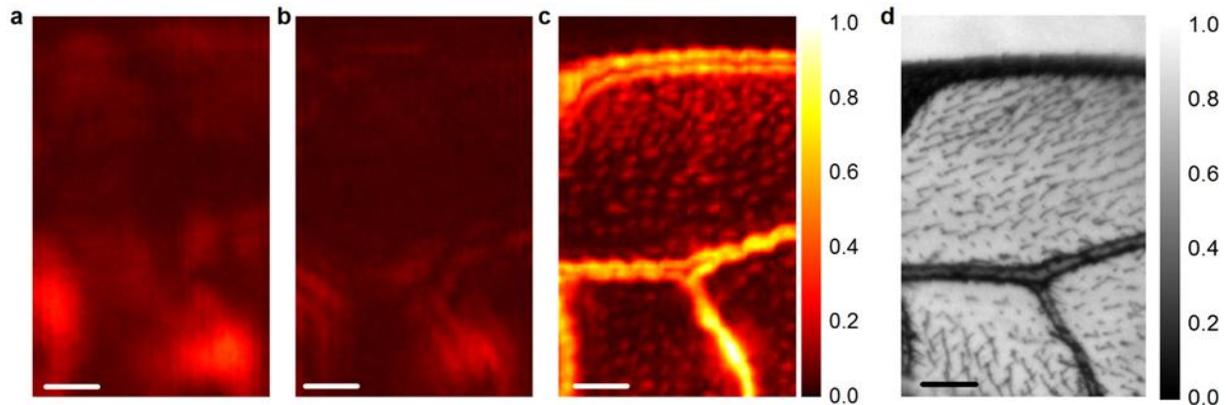

**Figure 4 | Photoacoustic image reconstructions of sweat bee wing.** The photoacoustic images created with different optical illuminations, scale bars are all 100 μm: **a**, acoustic resolution using uniform illumination (no scatterer), **b**, random optical speckle field from scatterer, **c**, wavefront optimized speckle focus. **d**, The optical image of the same section of wing.

In conclusion, we have proposed and demonstrated a technique to create an optical focus behind a scattering medium that enables improved resolution and SNR of photoacoustic imaging. The spatially varying feedback signal provided by the ultrasound transducer allowed the wavefront to evolve into a single speckle optical focus through mode competition. The method created an optical focal spot 5 times smaller than the acoustic focal spot size even though the optimization takes place on an extended absorber. The formation of this focus allowed for imaging absorbing samples at a higher resolution and SNR than possible without the scatterer. The technique is thus ideal for imaging through an opaque wall with resolution beyond that provided by the feedback mechanism. In this work, due to limitations imposed by a low laser repetition rate, the images presented were obtained by scanning the object while keeping the scatterer fixed. A straightforward extension would enable scanning of the laser beam and/or the acoustic transducer by continuous update of the wavefront[23]. Interestingly, the approach does not require use of the so-called memory effect[9,11] which would limit the applicability to weak or thin scatterers. The results presented here also suggest a path to increase the depth of optical resolution photoacoustic microscopy by providing high intensity optical foci deeper into tissue. This also opens up opportunities for 3D imaging applications in the diffusive regime beyond the sound wave diffraction-limited resolution, including medical imaging, photothermal therapy, neuroscience, and optogenetics.

## Methods
### Set-up

The experimental data are obtained using the set-up illustrated in Fig. 2a. A collimated 532 nm, 5 ns pulsed laser (Continuum Surelight I20, 20Hz repetition rate) illuminates a spatial light modulator (SLM) (Boulder Non-linear Systems, 512x512 pixels) for wavefront shaping. The SLM is imaged onto the back aperture of a long working distance microscope objective (5x, 0.14 NA). The wavefront focuses into a water tank and onto the surface of a glass diffuser. An absorbing sample ~8 mm behind the scatterer is mounted onto a 2D translation stage for scanning in the x and y dimensions. A 90 MHz (Olympus, V3512, 50 MHz bandwidth) focusing ultrasound transducer placed in the water detects the photoacoustic signal. After pre-amplification, an oscilloscope records, digitizes, and sends the detected signal to a computer for analysis.

To analyze the optical focus an Air Force resolution target is placed above the alpaca hair. After wavefront optimization the reflecting target is lowered into the focus to deflect the light to an imaging lens and CCD for analysis. The resolution target also provides calibration for the system magnification.

### Uniform and Speckle Illumination

Photoacoustic images generated using random speckle illumination (Fig. 3b and Fig. 4b) had the same speckle size as the optimized wavefront case. The speckle illumination is created by encoding a flat phase on the SLM. For uniform optical illumination the scatterer is removed from the optical beam path (Fig. 3a and Fig. 4a).

### Signal processing

During wavefront optimization low frequency signals are removed using a $2^{nd}$ order Butterworth digital high-pass filter with an 80 MHz cut-off frequency to limit the feedback area. In contrast, image reconstructions do not use the high-pass filter, but instead use the unfiltered signal. These images are reconstructed with the maximum peak to peak value from the waveform from each position.

**Acknowledgments**

We acknowledge support from the National Science Foundation award DGE-0801680. We also thank Youzhi Li for discussions.

**Author Contributions**

D.B.C, A.M.C., J.D.D., T.W.M., R.P. conceived the project and designed the research; D.B.C, A.M.C., and J.D.D. contributed equally to the design, execution, and analysis of the experiments. H.J. assisted with the execution of the experiment. D.B.C. performed the simulations; T.W.M. and R.P. supervised the project. All authors contributed to the preparation and revision of the manuscript.